 \documentclass[twocol]{ametsoc}

\usepackage{color}
\usepackage{amsmath}

\newcommand{\dy}{\partial}
\newcommand\ddy[2]{\frac{\dy#1}{\dy#2}}

\newcommand{\grad}{\nabla}

\newcommand{\eb}{\boldsymbol{e}}

\newcommand{\ub}{\boldsymbol{u}}

\definecolor{dark-green}{rgb}{0,0.5,0} 

\definecolor{ref-scheme}{rgb}{0,0.5,0} 
\definecolor{const-scheme}{rgb}{0.12,0.46,0.99} 
\definecolor{geom-scheme}{rgb}{1,0,0} 
\definecolor{constn2-scheme}{rgb}{0.54,0.81,0.94} 
\definecolor{geomn2-scheme}{rgb}{0.87,0.36,0.51} 
\definecolor{geomloc-scheme}{rgb}{0.41,0.16,0.38} 
\definecolor{geomn2loc-scheme}{rgb}{0.57,0.44,0.86} 

\journal{jpo}

\bibpunct{(}{)}{;}{a}{}{,}

\title{Implementation of a geometrically and energetically constrained
mesoscale eddy parameterization in an ocean circulation model}

\authors{J. Mak\correspondingauthor{Department of Physics, University of Oxford,
Oxford, OX1 3PU, United Kingdom}}

\affiliation{School of Mathematics and Maxwell Institute for Mathematical
Sciences, University of Edinburgh and\\
Department of Physics, University of Oxford}

\email{julian.c.l.mak@googlemail.com}

\extraauthor{J. R. Maddison}

\extraaffil{School of Mathematics and Maxwell Institute for Mathematical
Sciences, University of Edinburgh}

\extraauthor{D. P. Marshall}
\extraaffil{Department of Physics, University of Oxford}

\extraauthor{D. R. Munday}
\extraaffil{British Antarctic Survey, Cambridge}

\abstract{The global stratification and circulation of the ocean and their
sensitivities to changes in forcing depend crucially on the representation of
the mesoscale eddy field. Here, a geometrically informed and energetically
constrained parameterization framework for mesoscale eddies --- termed GEOMETRIC
--- is proposed and implemented in three-dimensional primitive equation channel
and sector models. The GEOMETRIC framework closes mesoscale eddy fluxes
according to the standard Gent--McWilliams scheme, but with the eddy transfer
coefficient constrained by the depth-integrated eddy energy field, provided
through a prognostic eddy energy budget evolving with the mean state. It is
found that coarse resolution calculations employing GEOMETRIC broadly reproduce
model sensitivities of the eddy permitting reference calculations in the
emergent circumpolar transport, meridional overturning circulation profile and
the depth-integrated eddy energy signature; in particular, eddy saturation
emerges in the sector configuration. Some differences arise, attributed here to
the simple prognostic eddy energy budget employed, to be improved upon in future
investigations. The GEOMETRIC framework thus proposes a shift in paradigm, from
a focus on how to close for eddy fluxes, to focusing on the representation of
eddy energetics.}

\begin{document}

\maketitle


%

\section{Introduction}

Accurate representation of the mesoscale eddy field and its feedback onto the
mean ocean state is one of the most pressing challenges for ocean modelling, in
particular in the ocean circulation models used for climate prediction, which
often lack explicit representation of the mesoscale eddy field. Over the past
two decades a widely adopted approach is due to \citet[][hereafter
GM]{GentMcWilliams90}. The GM scheme parameterizes eddies through both a
diffusion along neutral surfaces \citep{Redi82} and an eddy-induced circulation
that acts to flatten neutral density surfaces \citep{Gent-et-al95,
McDougallMcIntosh01}, thereby extracting available potential energy from the
mean state. The adoption of GM immediately resolved a number of known
deficiencies in ocean circulation models by removing the spurious diapycnal
water mass conversions that were prevalent in the existing eddy parameterization
schemes \citep{Danabasoglu-et-al94}.

A known deficiency of the GM eddy parameterization is the very different
response of the Southern Ocean circulation to changes in surface wind stress in
models with GM and explicit eddies. With a spatially constant eddy diffusivity,
the circumpolar transport increases with the strength of the surface wind
forcing, whereas little sensitivity is observed in the equivalent models with
explicit eddies \citep[e.g.,][]{Munday-et-al13, Farneti-et-al15}. This is known
as \emph{eddy saturation} \citep{HallbergGnanadesikan01} and was first predicted
on theoretical grounds by \citet{Straub93}. Eddy saturation is generally found
in models that partially resolve a mesoscale eddy field
\citep[e.g.,][]{HallbergGnanadesikan06, HoggBlundell06, Hogg-et-al08,
FarnetiDelworth10, Farneti-et-al10, MorrisonHogg13, Munday-et-al13,
HoggMunday14} but not in models where eddies are parameterized by the GM scheme
where the eddy transfer coefficient is constant in space and time
\citep[e.g.,][]{Munday-et-al13, Farneti-et-al15, Mak-et-al17}.

A further discrepancy between eddy permitting and coarse resolution models is
the reduced sensitivity of the time-mean residual meridional overturning
circulation to changing wind forcing obtained in eddy permitting models
\citep[e.g.,][]{Meredith-et-al12, ViebahnEden12, MorrisonHogg13, Munday-et-al13, 
HoggMunday14, Farneti-et-al15}. This is known as \emph{eddy compensation}
\citep{ViebahnEden12}. Eddy compensation is less well understood than eddy
saturation, depending in subtle ways on the vertical structure of the eddy
response to changes in surface forcing \citep[e.g.,][]{MorrisonHogg13}. The
response is further complicated by the fact that the residual meridional
overturning circulation is affected by, for example, bathymetric details
\citep[e.g.,][]{HoggMunday14, Ferrari-et-al16, DeLavergne-et-al17}. Generally,
it is found that eddy permitting calculations are strongly eddy saturated and
partially eddy compensated \citep[e.g.,][]{Munday-et-al13, Farneti-et-al15}.

In contrast, eddy saturation and compensation are not well represented in models
that parameterize eddies through GM with an eddy transfer coefficient that is
constant in space and time. Partial eddy saturation and compensation can be
obtained when the eddy transfer coefficient is allowed to vary in space and time
\citep[e.g.,][]{GentDanabasoglu11, HofmanMoralesMaqueda11, Farneti-et-al15}, due
to the nonlinear dependence of the eddy transfer coefficient on the mean density
gradients, in particular downstream of major bathymetric features. Numerous
papers have attempted to derive the functional dependence of the eddy transfer
coefficient on the ocean state as a function of space and time, from first
principles \citep[e.g.,][]{Treguier-et-al97, Visbeck-et-al97} and via diagnoses
of numerical simulations \citep[e.g.,][]{Ferreira-et-al05, Ferrari-et-al10,
BachmanFoxKemper13, Mak-et-al16b, Bachman-et-al17}. 

In \cite{EdenGreatbatch08} it was instead proposed that relating the eddy
transfer coefficient to the eddy kinetic energy through a mixing length argument
(see also \citealt{Cessi08}, \citealt{MarshallAdcroft10} and
\citealt{JansenHeld14}). This approach requires solving for the eddy kinetic
energy through a prognostic eddy energy budget. More recently,
\cite{Marshall-et-al12} have developed a new framework, here termed
``GEOMETRIC'', in which the inferred GM eddy transfer coefficient is entirely
determined by the total eddy energy, the stratification, and an unknown
non-dimensional parameter that is bounded in magnitude by unity. The predicted
eddy transfer coefficient broadly agrees with that diagnosed in eddy resolving
calculations \citep{Bachman-et-al17}. Moreover, the GEOMETRIC eddy transfer
coefficient leads to eddy saturation when implemented in an idealized
two-dimensional model of the Antarctic Circumpolar Current \citep{Mak-et-al17}.

The aims of this work are to:
\begin{enumerate}
  \item implement GEOMETRIC in a three-dimensional ocean circulation model;
  \item diagnose the extent to which GEOMETRIC reproduces eddy saturation and
  eddy compensation as obtained in the eddy permitting calculations;
  \item explore the spatial variations of the eddy energy in the eddy permitting
  calculations and the extent to which these are reproduced with GEOMETRIC.
\end{enumerate}

The article proceeds as follows. Section~\ref{sec:geom} outlines the GEOMETRIC
approach and section~\ref{sec:eddy-ene} discusses the associated parameterized
eddy energy budget. Implementation details relating to the parameterization
schemes considered in this article are given in Section~\ref{sec:param} Results
from idealized channel and sector configurations are detailed in
Section~\ref{sec:channel} and \ref{sec:sector} respectively, with the model
numerics described within the sections. The article summarizes and concludes in
Section~\ref{sec:conc}, and discusses further implementation challenges and
research directions.


\section{GEOMETRIC}\label{sec:geom}

GEOMETRIC (``Geometry of Ocean Mesoscale Eddies and Their Rectified Impact on
Climate'') represents a framework for parameterizing mesoscale eddies that
preserves the underlying symmetries and conservation laws in the un-averaged
equations of motion. GEOMETRIC was originally derived under the
quasi-geostrophic approximation \citep{Marshall-et-al12} although elements of
the framework generalize to the thickness-weighted averaged primitive equations
\citep{MaddisonMarshall13}. 

There are two fundamental ingredients in GEOMETRIC: 
\begin{enumerate}
  \item representation of the eddy-mean flow interaction through an eddy stress
  tensor, which can be bounded in terms of the total eddy energy
  \citep{Marshall-et-al12}; 

  \item solution of a consistent eddy energy equation
  \citep[cf.][]{EdenGreatbatch08, Cessi08, MarshallAdcroft10}. 
\end{enumerate} 
Crucially, given knowledge of the total eddy energy and mean stratification, all
of the remaining unknowns are dimensionless, i.e., there is no freedom to
specify dimensional quantities such as eddy length scales or eddy diffusivities. 

In the simplest limit, in which the lateral eddy Reynolds stresses are
neglected, GEOMETRIC reduces to GM, with the eddy transfer coefficient given by  
\begin{equation}\label{eq:rawGEOM}
  \kappa_{\rm gm} = \alpha \, E \, \frac{N}{M^2}, 
\end{equation}
where $E$ is the total eddy energy, $N = (\dy\overline{b} / \dy z)^{1/2}$ is the
buoyancy frequency, $M^2 = |\grad_H \overline{b} |$ is the magnitude of the
lateral buoyancy gradient, $b$ is buoyancy and $\grad_H$ is the horizontal
gradient operator. Here the overbar represents a time filter applied at fixed
height. An equivalent form of this was given in \citet{Jansen-et-al15b},
obtained through combining a mixing length argument with those of
\citet{LarichevHeld95}, but with the eddy kinetic energy in place of the total
eddy energy. 

Once the eddy energy field is known, the only freedom then is in the
specification of the non-dimensional eddy efficiency parameter $\alpha$,
satisfying $|\alpha| \leq 1$ in the quasi-geostrophic limit
\citep{Marshall-et-al12}. This $\alpha$ parameter can be diagnosed from eddy
permitting simulations: results from wind-driven gyre calculations in a
quasi-geostrophic model \citep{Marshall-et-al12} and nonlinear Eady spindown
calculations in a primitive equation model \citep{Bachman-et-al17} suggest that
typically $\alpha = O(10^{-1})$. 

The efficacy of GEOMETRIC has been established through three proofs of concept: 
\begin{itemize}
  \item in the linear \citet{Eady49} model of baroclinic instability, an
  analytical test case, GEOMETRIC produces the correct dimensional energy growth
  rate \citet{Marshall-et-al12};
  
  \item in the fully-turbulent nonlinear Eady spin-down problem, as simulated by
  \citet{Bachman-et-al17}, the eddy transfer coefficient predicted by GEOMETRIC,
  (\ref{eq:rawGEOM}), gives good agreement with those diagnosed from the
  numerical calculations, across four orders of magnitude of the eddy
  transfer coefficient;
  
  \item when applied to a two-dimensional model of the Antarctic Circumpolar
  Current with a domain-integrated eddy energy budget \citep{Mak-et-al17},
  GEOMETRIC produces eddy saturation, i.e., a circumpolar volume transport that
  is insensitive to the surface wind stress, due to an interplay with the zonal
  momentum budget and eddy energy budget \citep{Marshall-et-al17}, the essential
  components of which are preserved by GEOMETRIC.
\end{itemize}


\section{Eddy energy equation}\label{sec:eddy-ene}

The outstanding challenge is then to solve for the eddy energy field. Solution
of a prognostic equation for the kinetic eddy energy in three dimensions has
been attempted by \citet{EdenGreatbatch08}. In GEOMETRIC, the total eddy energy
is required. In this paper, it is proposed that the \emph{depth-integrated} eddy
energy is solved for, as this offers a number of advantages: (i) the conceptual
and logistical simplicity of working in two rather than three dimensions; (ii)
avoidance of division by zero in \eqref{eq:rawGEOM} when the isopycnals are
flat; (iii) retention of desirable properties of GM such as the
positive-definite sink of available potential energy \citep{GentMcWilliams90,
Gent-et-al95} that are instrumental in its robustness.

The consequence is that the eddy diffusivity is energetically constrained in the
vertical integral only. Specifically, suppose the eddy transfer coefficient
varies in the vertical according to 
\begin{equation}
  \kappa_{\rm gm}(z) = \kappa_0 \Gamma (z) 
\end{equation}
where $\Gamma (z)$ is a prescribed dimensionless structure function,
\citep[e.g., $\Gamma (z) = N(z)^2/N_{\rm ref}^2$,][]{Ferreira-et-al05}. Then the
proposed eddy transfer coefficient is
\begin{equation}\label{eq:GEOMlockappa}
  \kappa_{\rm gm} = \alpha \frac{\int E\; \mathrm{d}z}{\int (\Gamma M^2 / N)\; \mathrm{d}z} \, \Gamma (z),  
\end{equation}
which is to be coupled to a parameterized budget for the depth-integrated eddy
energy, $\int E \, \mathrm{d}z$.

Rather than derive a depth-integrated eddy energy budget from first principles,
which contains terms that are unknown, the following heuristic approach is
taken. The primary source of eddy energy is baroclinic instability
\citep{Charney48, Eady49}, and must be incorporated in a manner that is
consistent with the loss of mean energy due to the slumping of density surfaces
as represented by GM. 

In observations, the eddy energy is observed to propagate westward, at roughly
the intrinsic long Rossby phase speed \citep{Chelton-et-al07, Chelton-et-al11},
with an additional advective contribution that is adequately modelled by the
depth-mean flow \citep{KlockerMarshall14}. In this paper, the contribution from
the intrinsic long Rossby phase speed is not included, while recognising that it
will be required in an eventual implementation of GEOMETRIC in a global
circulation model. Previous studies indicate that the lateral
redistribution of eddy energy is not required to obtain eddy saturation with
GEOMETRIC \citep{Marshall-et-al17, Mak-et-al17}, but it will likely affect the
detailed response.

A Laplacian diffusion of eddy energy is incorporated following
\cite{EdenGreatbatch08}. While included as a stabiliser, there are indications
that the use of a Laplacian diffusion corresponds to the divergence of the mean
energy flux in an $f$-plane barotropic model of turbulence \citep{Grooms15}.

The dissipation of eddy energy is complicated, involving a myriad of processes.
These include: bottom drag \citep[e.g.,][]{Sen-et-al08}; lee wave radiation from
the sea floor \citep[e.g.,][]{NaveiraGarabato-et-al04, NikurashinFerrari11,
Melet-et-al15}; western boundary processes \citep{Zhai-et-al10}; and loss of
balance \citep[e.g.,][]{Molemaker-et-al05}. Moreover, the eddy energy
dissipation through these various processes will critically depend on the
partition between eddy kinetic and eddy potential energy, and the vertical
structure of the eddy kinetic energy \citep{Jansen-et-al15b, KongJansen17}. Each
of these required detailed investigation. Instead, a simple approach is followed
here, representing eddy energy dissipation through a linear damping at a rate
$\lambda$, recognising that $\lambda$ parameterizes all of the physics outlined
above.

To summarize, the proposed parameterized eddy energy budget is
\begin{align}\label{eq:GEOMloc-e}
  \ddy{}{t}&\int E\; \mathrm{d}z
    + \grad_H \cdot \left( \left(\widetilde{\ub}^z - c\, \eb_x\right) \int E\; \mathrm{d}z \right) \nonumber \\
  &= \int \kappa_{\rm gm} \frac{M^4}{N^2}\; \mathrm{d}z
    - \lambda \int E\; \mathrm{d}z + \eta_E\grad^2_H  \int E\; \mathrm{d}z,
\end{align}
where: $\widetilde{\ub}^z$ is the depth-averaged flow; $\eb_x$ is the unit
vector in the longitudinal direction, and $c$ is the intrinsic long Rossby phase
speed which varies with latitude \citep{Chelton-et-al07, Chelton-et-al11};
$\kappa_{\rm gm} M^4 / N^2$ is the eddy energy source, equal to the release of
mean potential energy by slumping of density surfaces via GM; $\lambda$ is the
linear damping coefficient for the eddy energy that could in principle be a
function of space and time; $\eta_E$ is the coefficient of the Laplacian
diffusion of the depth-integrated eddy energy.


\section{Experimental design}\label{sec:param}

As a first step, the following simplifications are made: $\kappa_{\rm gm}$ is
taken to be vertically constant (so $\Gamma(z) \equiv 1$); $c$ is set to zero;
$\lambda$ is a control parameter that is a constant in space time. Three sets of
experiments are considered, as described by the following.

\subsubsection{GEOM$_{\rm loc}$}

The first set of experiments employ GEOMETRIC locally in latitude and longitude,
as detailed in Section~\ref{sec:geom} and \ref{sec:eddy-ene}, with the eddy
transfer coefficient computed as \eqref{eq:GEOMlockappa}, coupled to the
parameterized eddy energy budget \eqref{eq:GEOMloc-e}. The \mbox{GEOM$_{\rm
loc}$} scheme is implemented wholly within the \verb|GM/Redi| package within
MITgcm \citep{Marshall-et-al97a, Marshall-et-al97b}. First, all spatial
derivatives of the mean density field are passed through a five point smoother,
and $\kappa_{\rm gm}$ is calculated according to \eqref{eq:GEOMlockappa} with
the smoothed $M^2/N$ (the vertical integral of $M^2 / N$ is bounded below by a
small number to prevent division by zero). Then, if desired as a precaution to
prevent possible large eddy induced velocities, $\kappa_{\rm gm}$ may be capped
below and above by some $\kappa_{\min}$ and $\kappa_{\max}$, and the GM/Redi
tensor is formed and passed through a slope tapering/clipping scheme. With the
resulting $\kappa_{\rm gm}$, the eddy energy budget \eqref{eq:GEOMloc-e},
discretized in space by a centered second-order differencing, is time stepped
with a third order Adams--Bashforth scheme (started with forward and second
order Adams--Bashforth steps), again with the smoothed $M^2/N$.

In this work, $\eta_E = 2000\ \mathrm{m}^2\ \mathrm{s}^{-1}$, $\kappa_{\min} =
50\ \mathrm{m}^2\ \mathrm{s}^{-1}$, $\kappa_{\max} = 15000\ \mathrm{m}^2\
\mathrm{s}^{-1}$, and the \verb|gkw91| slope tapering scheme
\citep{Gerdes-et-al91} was chosen with the maximum slope parameter set to be $5
\times 10^{-3}$.

\subsubsection{GEOM$_{\rm int}$}

In the second set of experiments, \mbox{GEOM$_{\rm int}$}, as previously
considered in \cite{Mak-et-al17}, the eddy energy budget is integrated in space.
With $x$ as longitude and $y$ as latitude, the eddy transfer coefficient is
calculated as
\begin{equation}\label{eq:GEOMkappa}
  \kappa_{\rm gm} = \alpha \frac{\iiint E\; \mathrm{d}z\; \mathrm{d}x\; \mathrm{d}y}{\iiint (M^2/N)\; \mathrm{d}z\; \mathrm{d}x\; \mathrm{d}y},
\end{equation}
and is coupled to the parameterized eddy energy budget given by
\begin{align}\label{eq:GEOM-e}
  \frac{\mathrm{d}}{\mathrm{d}t}\iiint E\; \mathrm{d}z\; \mathrm{d}x\; \mathrm{d}y 
  &= \iiint \kappa_{\rm gm} \frac{M^4}{N^2}\; \mathrm{d}z\; \mathrm{d}x\; \mathrm{d}y \nonumber\\
    &- \lambda \iiint E\; \mathrm{d}z\; \mathrm{d}x\; \mathrm{d}y,
\end{align}
where there are no longer advective contributions to the eddy energy tendency,
and the Laplacian diffusion of eddy energy has been removed. The eddy transfer
coefficient \eqref{eq:GEOMkappa} is now a constant in space but may vary in
time, and the eddy energy budget \eqref{eq:GEOM-e} becomes an ordinary
differential equation. The \mbox{GEOM$_{\rm int}$} scheme was also implemented
wholly within the \verb|GM/Redi| package in MITgcm following analogous steps,
except the eddy energy budget \eqref{eq:GEOM-e} is time-stepped by a backward
Euler scheme for numerical stability. The same $\kappa_{\min}$, $\kappa_{\max}$
and slope tapering scheme as \mbox{GEOM$_{\rm loc}$} were used.

\subsubsection{CONST}

Finally, a control case with the standard GM scheme and a constant, prescribed
eddy transfer coefficient
\begin{equation}\label{eq:CONSTkappa}
  \kappa_{\rm gm} = \kappa_0 
\end{equation}
is considered. The emergent parameterized eddy energy does not affect any of the
resulting dynamics, and the routines for time stepping the parameterized eddy
energy budget are bypassed.

\vspace*{3mm}

The coarse resolution calculations \mbox{GEOM$_{\rm int}$}, \mbox{GEOM$_{\rm
loc}$} and \mbox{CONST} are compared to calculations from eddy permitting
reference calculations (REF). To assess the performance of the parameterization
variants, various diagnoses of the resulting time-averaged data are presented;
unless otherwise stated, all subsequent figures and statements refer to the
time-averaged data. No mixed layer schemes are employed in the calculations.


The theory behind GEOMETRIC applies to the GM eddy transfer coefficient and not
to the enhanced eddy diffusion of tracers along isopycnals
\citep[e.g.,][]{Redi82}. While GM and Redi diffusion are often implemented in
the GM/Redi tensor together \citep[e.g.,][]{Griffies98, Griffies-et-al98}, the
corresponding coefficients need not be the same. In all calculations presented
here, the Redi diffusion coefficient is prescribed to be $\kappa_{\tiny
\mbox{redi}} = 200\ \mathrm{m}^2\ \mathrm{s}^{-1}$, and the GM eddy transfer
coefficient follows the prescription of \mbox{GEOM$_{\rm int}$},
\mbox{GEOM$_{\rm loc}$} or \mbox{CONST} as appropriate. The consequences of the
parameterization choices as well as the simplifications made for this work are
discussed at the end of the article.


\section{Channel configuration}\label{sec:channel}


\subsection{Setup and diagnostics}

As an extension of the $f$-plane, zonally-averaged channel model of the
Antarctic Circumpolar Current presented in \cite{Mak-et-al17}, an idealized
channel configuration on a $\beta$-plane is considered in MITgcm. The
configuration is essentially a shorter version of the channel configuration
reported in \cite{Munday-et-al15} and \cite{Marshall-et-al17}, with no
continental barriers. The domain is $4000\ \mathrm{km}$ long, $2000\
\mathrm{km}$ wide and with a maximum depth of $3000\ \mathrm{m}$. The model
employs a linear equation of state with temperature only, and with an implicit free
surface. A ridge with height of $1500\ \mathrm{m}$ and width $800\ \mathrm{km}$
blocks $f/H$ contours and allows for the topographic form stress to balance the
surface wind stress \cite[e.g.,][]{MunkPalmen51, JohnsonBryden89}. 

An idealized zonal wind stress of the form
\begin{equation}\label{eq:wind-channel}
  \tau_s = \frac{\tau_0}{2} \left(1 + \cos\left(\frac{2\pi y}{L_y}\right)\right)
\end{equation}
is imposed, where $L_y$ is the meridional width of the channel, and $\tau_0$ is
the peak wind stress. The temperature is restored to the linear profile
\begin{equation}\label{eq:temp-surf-channel}
  T = \left(\frac{y + L_y / 2}{L_y}\right)\Delta T
\end{equation}
with $\Delta T = 15\ \mathrm{K}$ on a time-scale of $10\ \mathrm{days}$ over the
top cell of height $10\ \mathrm{m}$. The vertical temperature diffusion has
magnitude $\kappa_d = 10^{-5}\ \mathrm{m}^2\ \mathrm{s}^{-1}$, except in a
tapered sponge region to the north of width $150\ \mathrm{km}$ where the
vertical temperature diffusion is increased sinusoidally to $\kappa_d = 5\times
10^{-3}\ \mathrm{m}^2\ \mathrm{s}^{-1}$ to maintain a non-trivial stratification
and energize the eddies \citep[e.g.][]{Hogg10, Munday-et-al15}. A linear bottom
drag with coefficient $r$ is applied in the deepest level above the bathymetry.
The vertical domain is discretized with $30$ uneven vertical levels, thinnest of
$10\ \mathrm{m}$ at the top, down to the thickest of $250\ \mathrm{m}$, and with
a partial cell representation of the bathymetry. A staggered baroclinic time
stepping scheme was employed. See \cite{Munday-et-al15} and
\cite{Marshall-et-al17} for further model details.

For the eddy permitting reference calculations (REF), the horizontal grid
spacing is uniform at $10\ \mathrm{km}$. A control simulation with control peak
wind stress $\tau_0 = \tau_c = 0.2\ \mathrm{N}\ \mathrm{m}^{-2}$ and control
bottom drag coefficient $r = r_c = 1.1\times10^{-3}\ \mathrm{m}\
\mathrm{s}^{-1}$ is carried out for $400$ model ``years'' (360 days), from which
perturbation experiments at varying $\tau_0$ and $r$ were carried out for a
further $200$ model years, before averages were taken for a further $20$ model
years. The eddy permitting calculation employs the full Leith viscosity
\citep[e.g.,][]{FoxKemperMenemenlis08} with a coefficient of $2$.

For the coarse resolution models, the horizontal grid spacing is mostly at $100\
\mathrm{km}$, except at the northern boundary where the grid spacing is $50\
\mathrm{km}$ so as to have at least three grid points over the northern sponge
region. A control \mbox{GEOM$_{\rm int}$} calculation with $\tau_c$ and $\lambda
= \lambda_c = 10^{-7}\ \mathrm{s}^{-1}$ (consistent with observation-constrained
estimates in the Southern Ocean, Marshall \& Zhai, pers. comm.; see also
\citealt{Melet-et-al15}) is first carried out over $500$ model years.
Perturbation experiments in \mbox{GEOM$_{\rm int}$}, \mbox{GEOM$_{\rm loc}$} and
\mbox{CONST} at varying $\tau_0$ and $\lambda$ (rather than $r$) are then
restarted and carried out for a further $300$ model years, and averages are
taken for a further $200$ model years. The coarse resolution calculations
employ a harmonic friction in the momentum equation that forces the grid scale
Reynolds number to be $0.0075$. The calculations have $\alpha$ and $\kappa_0$
tuned so that the circumpolar transport at the control parameter values matches
the control calculation of REF, after which they are fixed for the perturbation
experiments. 

Note that in REF, the control parameters are $\tau_0$ and $r$, while in the
coarse resolution calculations, the control parameters $\tau_0$ and $\lambda$,
and $r$ is fixed. The relevant parameter values are documented in
Table~\ref{tbn:channel-params}.

\begin{table*}[tbp]
  \begin{center}
  {\small
    \begin{tabular}{|l|c|c|}
      \hline
      Parameter & Value & units\\
      \hline
      $\tau_0$ & 
        0.00, 0.05, 0.10, 0.15, \underline{0.20}, 0.25, 0.30, 0.40, 0.60, 0.80, 1.00 &
        $\mathrm{N}\ \mathrm{m}^{-2}$\\
      \hline
      $r$ & 
        0.55, 0.66, 0.77, 0.88, 0.99, \underline{1.10}, 2.20, 3.30, 4.40, 5.50 &
        $10^{-3}\ \mathrm{m}\ \mathrm{s}^{-1}$\\
      \hline
      $\lambda$ & 
        0.95, \underline{1.00}, 1.10, 1.20, 1.30, 1,40, 1.50 &
        $10^{-7}\ \mathrm{s}^{-1}$\\
      \hline
      $\kappa_0$ &
        1500 (\mbox{CONST}) &
        $\mathrm{m}^2\ \mathrm{s}^{-1}$\\
      \hline
      $\alpha$ &
        0.04 (\mbox{GEOM$_{\rm int}$}), 0.042 (\mbox{GEOM$_{\rm loc}$}) &
        ---\\
      \hline
    \end{tabular}
  }
  \end{center}
  \caption{Parameter values that are employed for the channel experiments. The
  values underlined are designated the control simulation.}
  \label{tbn:channel-params}
\end{table*}

Several diagnostics are computed to compare mean properties of the
parameterization variants \mbox{GEOM$_{\rm int}$}, \mbox{GEOM$_{\rm loc}$} and
\mbox{CONST}, against the reference calculation REF. The total circumpolar
transport given by
\begin{equation}\label{eq:transtotal}
  T_{\scriptsize \mbox{tot}} = \frac{1}{L_x} \int\left(
    \iint \overline{u}\; \mathrm{d}y\; \mathrm{d}z
  \right)\; \mathrm{d}x
\end{equation}
where $L_x$ is the length of the circumpolar channel, and $\overline{(\cdot)}$
denotes a time filter performed at fixed height. Another is the thermal wind
transport, given by
\begin{equation}\label{eq:transtherm}
  T_{\scriptsize \mbox{therm}} = \frac{1}{L_x} \int\left(
    \iint \overline{u}_{\tiny \mbox{therm}}\; \mathrm{d}y\; \mathrm{d}z
  \right)\; \mathrm{d}x
\end{equation}
the thermal wind velocity is given by
\begin{equation}
  \overline{u}_{\scriptsize \mbox{therm}}
    = \int\frac{g \rho_0}{f_0 + \beta y} \ddy{\overline{\rho}}{y}\; \mathrm{d}z,
\end{equation}
assuming that $\overline{u}_{\tiny \mbox{therm}}(z = -H) = 0$, with
$\overline{\rho}$ obtained from the temperature via the linear equation of
state. Then, the complement transport to thermal wind transport is defined to be
\begin{equation}\label{eq:transcom}
  T_{\scriptsize \mbox{com}} = T_{\scriptsize \mbox{tot}} - T_{\scriptsize \mbox{therm}}.
\end{equation}
Note that the definition of the transport decomposition employed here differs
from that employed in \cite{Munday-et-al15}. There, the bottom flow transport is
defined to be equal to the bottom flow multiplied by the depth (thus sensitive
to bottom flow details), and the baroclinic transport then is the remaining
component of the total transport. It should be noted that the diagnosed values
of the baroclinic transport as defined in \cite{Munday-et-al15} and the thermal
wind transport defined in \eqref{eq:transtherm} differ only very slightly in
these channel experiments. Finally, following the definition of
\cite{Gnanadesikan99} (see also \citealt{AbernatheyCessi14}), a thermocline
location is obtained by computing
\begin{equation}\label{eq:therm-depth} 
  z_{\scriptsize \mbox{therm}} = 2 \cfrac{\int_{-H}^0 z[\overline{T} - \overline{T}(z=-H)]\; \mathrm{d}z}{\int_{-H}^0 [\overline{T} - \overline{T}(z=-H)]\; \mathrm{d}z},
\end{equation}
essentially a center-of-mass calculation for the vertical co-ordinate $z$, and
this quantity is averaged over the northern sponge region where the thermocline
is deepest. This provides a measure of the model stratification, with a deeper
thermocline (more negative $z_{\scriptsize \mbox{therm}}$) expected to correlate
with increased thermal wind transport.



\subsection{Results}

\begin{figure}
  \includegraphics[width=\linewidth]{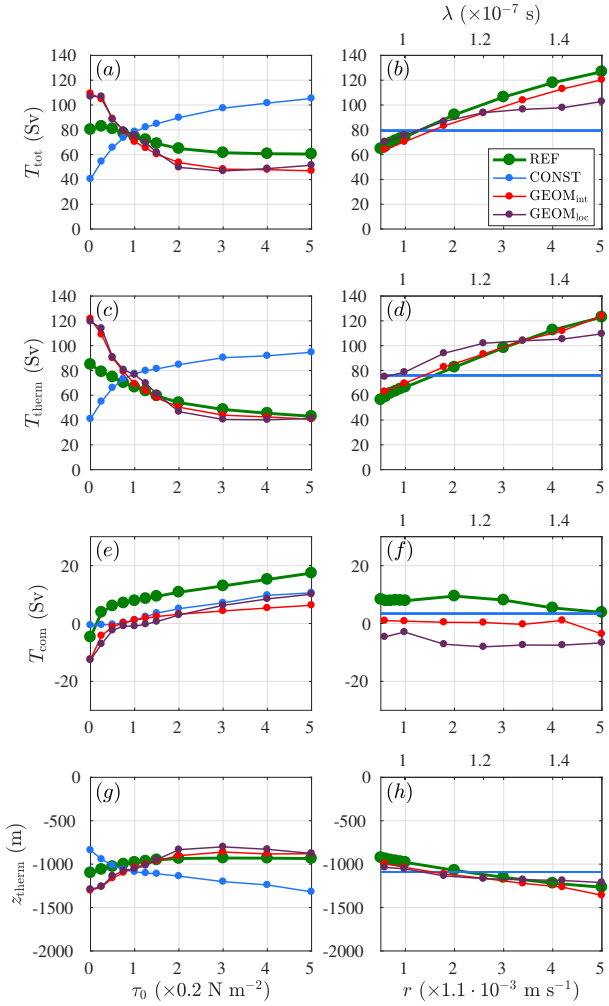}
  \caption{Diagnosed transports (in units of Sv) and thermocline location (in
  units of $\mathrm{m}$) in the channel model, for varying wind and varying
  dissipation. Showing: ($a,b$) total transport; ($c,d$) thermal wind transport;
  ($e,f$) complement transport to thermal wind; ($g,h$) thermocline location.
  While the REF calculations vary $r$, in the bottom axis, it is $\lambda$ that
  is varied in the \mbox{GEOM$_{\rm int}$} and \mbox{GEOM$_{\rm loc}$}
  calculations, displayed in the top axis ($\lambda$ does not affect
  \mbox{CONST}).}
  \label{fig:channel_tave_vary_diag}
\end{figure}

The diagnosed results are presented in Figure~\ref{fig:channel_tave_vary_diag}.
As a summary, in this channel set up, the total transport of REF decreases with
increasing wind, and increases with increased linear bottom drag. The total
transport is composed principally of transport due to thermal wind. The
complement component to the thermal wind transport increases with increased
wind, and decreases slightly with increased bottom linear drag. The changes in
the thermal wind transport are reflected in the resulting $z_{\scriptsize
\mbox{therm}}$, where a deeper thermocline corresponds to a larger spatial
extent of the thermal wind. With this in mind, the \mbox{CONST} calculations
display the opposite sensitivity to REF in terms of the dependence of the total
circumpolar transport, thermal wind transport and the resulting thermocline
location on the peak wind stress magnitude. On the other hand, both
\mbox{GEOM$_{\rm int}$} and \mbox{GEOM$_{\rm loc}$} capture the changes in
thermal wind transport and thermocline location with increasing wind stress.
While there is a degree of tuning for the range of $\lambda$ exployed here, the
\mbox{GEOM$_{\rm int}$} and \mbox{GEOM$_{\rm loc}$} simulations at these values
of $\lambda$ results in similar trends to REF for the total and thermal wind
transport.

\subsubsection{Varying wind experiments}

First, it is interesting to note that, even in REF, the total transport
decreases with increasing wind, and the transport is non-zero at zero wind. The
latter is due to the northern sponge region with enhanced vertical temperature
diffusivity, which acts to maintain a stratification at depth and, together with
surface restoring of temperature, results in tilting isopycnals and thus a
thermal wind transport \citep[e.g.,][]{MorrisonHogg13}. In this model, the model
thermocline becomes shallower with increasing wind. As a result, the
geostrophic flow occupies a smaller volume even though the peak geostrophic flow
speed may be larger, thus resulting in a smaller integrated thermal wind
transport. The decreased thermocline depth with increasing wind is likely due to
the choice of imposing the northern sponge region condition; such behavior is
not observed when a fully dynamical basin sets the northern channel
stratification (as in the sector profile in the next section) or when the
northern boundary temperature is relaxed to a prescribed profile (as in, e.g.,
\citealt{AbernatheyCessi14}, where they employ instead a flux boundary condition
at the ocean surface).

\begin{figure}
  \includegraphics[width=\linewidth]{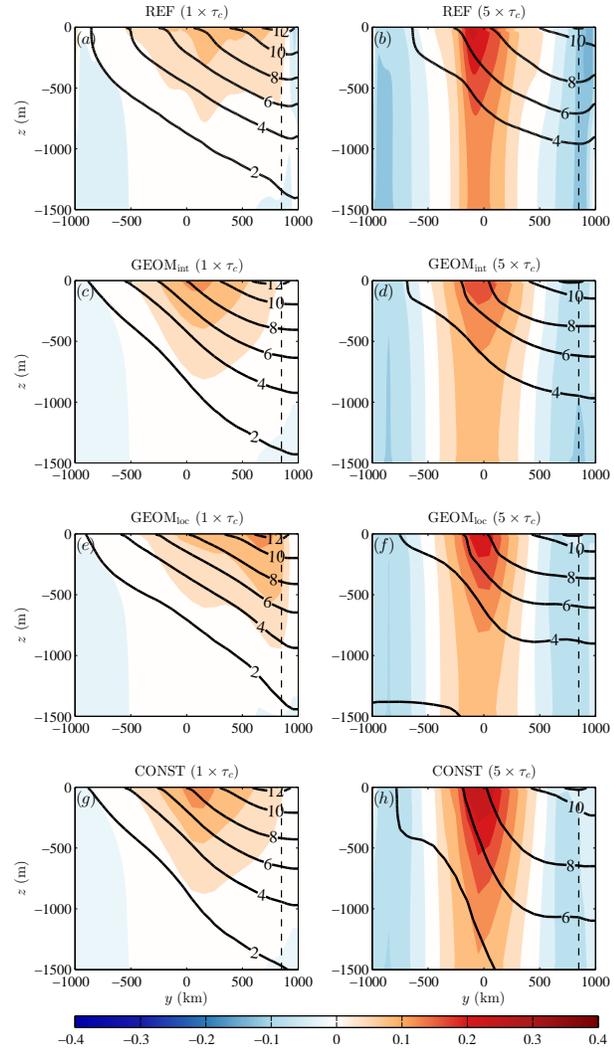}
  \caption{Zonally averaged zonal velocity (shaded, in units of $\mathrm{m}\
  \mathrm{s}^{-1}$) and zonally-averaged temperature (contours, in ${}^\circ
  \mathrm{C}$) over the top half (1500 $\mathrm{m}$) of the domain, for control
  wind strength and five times the control peak wind strength. Showing: ($a,b$)
  REF; ($c,d$) \mbox{GEOM$_{\rm int}$}; ($e,f$) \mbox{GEOM$_{\rm loc}$}; ($g,h$)
  \mbox{CONST}. The black dashed line in all panels denotes the boundary between
  the interior and the northern sponge region.}
  \label{fig:channel_tave_ref_vary_wind}
\end{figure}

Despite the perhaps unexpected sensitivity to changing wind forcing in
\mbox{REF}, it is encouraging to see that both the \mbox{GEOM$_{\rm int}$} and
\mbox{GEOM$_{\rm loc}$} are able to reproduce the analogous sensitivities,
particularly in the thermal wind transport and thermocline location diagnostic.
In contrast, the standard \mbox{CONST} variant displays opposite sensitivity in
the transport and thermocline location.
Figure~\ref{fig:channel_tave_ref_vary_wind} shows the emergent zonally averaged
temperature profile and zonal flow of the eddy permitting calculation and coarse
resolution calculations. \mbox{GEOM$_{\rm int}$} and \mbox{GEOM$_{\rm loc}$} are
able to capture the changes in the stratification displayed by REF. An
examination of the absolute difference in zonally-averaged zonal velocity (not
shown) shows the largest discrepancies lie within the high diffusivity sponge
region, where the coarse resolution calculations generally under-estimate the
zonal flow.

\subsubsection{Varying dissipation experiments}

For increasing bottom drag, the total transport of \mbox{REF} increases,
consistent with the results of \cite{Marshall-et-al17}. The rationale is that
increased dissipation reduces the effective eddy induced overturning that
counteracts the Eulerian overturning, resulting in steeper isopycnals. This
leads to increased thermal wind transport, and is consistent with the
diagnostics displayed in Figure~\ref{fig:channel_tave_vary_diag}($b,d$). This
feature of increased thermal wind transport is reproduced by the
\mbox{GEOM$_{\rm int}$} and \mbox{GEOM$_{\rm loc}$} calculations, and is
consistent with the findings of \cite{Mak-et-al17}.


\subsubsection{Other emergent quantities}

\begin{figure}
  \includegraphics[width=\linewidth]{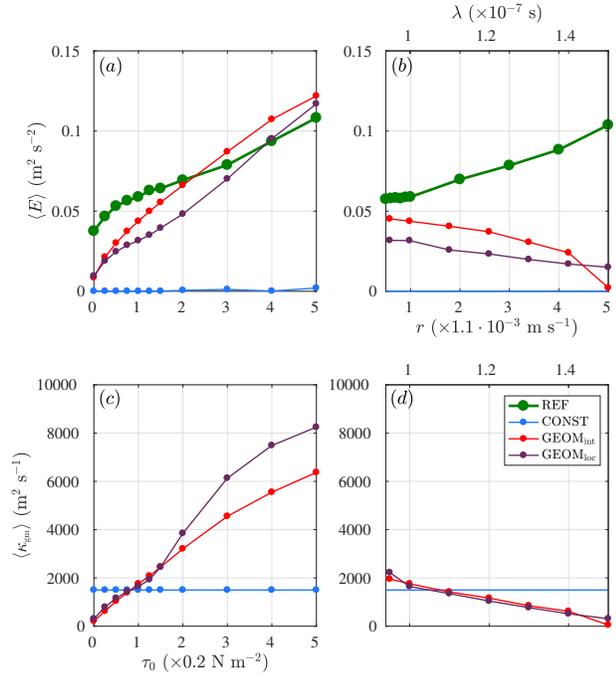}
  \caption{Diagnosed outputs relating to the paramaeterisation variants for the
  channel model, for varying wind ($a,c$) and varying dissipation ($b,d$),
  showing: ($a,b$) domain-averaged eddy energy (in units of $\mathrm{m}^2\
  \mathrm{s}^{-2}$); ($c,d$) domain-averaged GM coefficient $\kappa_{\rm gm}$
  for parameterized models (in units of $\mathrm{m}^2\ \mathrm{s}^{-1}$). There
  are no diagnosed $\kappa_{\rm gm}$ values for REF in panels ($c,d$).}
  \label{fig:channel_tave_vary_diag_eke_kappa}
\end{figure}

The emergent eddy energy level and GM eddy transfer coefficient $\kappa_{\rm
gm}$ have also been diagnosed. Figure~\ref{fig:channel_tave_vary_diag_eke_kappa}
shows the domain-averaged eddy energy $\langle E\rangle$ and domain-averaged GM
eddy transfer coefficient $\langle \kappa_{\rm gm} \rangle$. While the
\mbox{GEOM$_{\rm int}$} and \mbox{GEOM$_{\rm loc}$} calculations have a value of
the total eddy energy from the parameterized eddy energy budget, the
depth-integrated value of the total eddy energy\footnote{More precisely, the
specific total eddy energy, with units of $\mathrm{m}^2\ \mathrm{s}^{-2}$.} for
REF and \mbox{CONST} is calculated by diagnosing the sum of the depth-integrated
(specific) eddy kinetic energy
\begin{equation}\label{eq:EKE}
  \int\mbox{EKE}\ \mathrm{d}z = 
  \frac{1}{2}\int_{-H(x,y)}^0\left(\overline{u'u'} +
   \overline{v'v'}\right)\; \mathrm{d}z
\end{equation}
and depth-integrated (specific) eddy potential energy, as (see, for example
Ch.~3, of \citealt{Vallis-GFD})
\begin{equation}\label{eq:EPE}
  \int_{-H(x,y)}^0\mbox{EPE}\ \mathrm{d}z =
  \frac{1}{2}\int_{\rho_b}^{\rho_t} \frac{g}{\rho_0} \overline{z'z'}\; \mathrm{d}\rho.
\end{equation}
Here, $\ub = \overline{\ub} + \ub'$ and $z = \widehat{z} + z'$, where the time
filter on the latter is to be carried out in density co-ordinates, so $z'$ is
the deviation from the mean isopycnal height. The thickness-weighted averaging
is carried out here with the \verb|layers| package in MITgcm
\citep[e.g.,][]{Abernathey-et-al11}. For this channel configuration with a
linear equation of state for temperature, the calculation is carried out in
temperature co-ordinates, with temperature referenced to the top model level at
the surface, with binning over 81 discrete layers between $-$4${}^\circ\
\mathrm{C}$ and 16${}^\circ\ \mathrm{C}$, equally spaced at 0.25${}^\circ\
\mathrm{C}$. In the eddy permitting calculations, it is the EPE contributions
that dominate, accounting for around 90\% of the total eddy energy; at the
highest wind forcing, EKE accounts for about 12\% of the total eddy energy, and
decreases to about 5\% for the largest value of linear bottom drag coefficient
employed.

\begin{figure*}
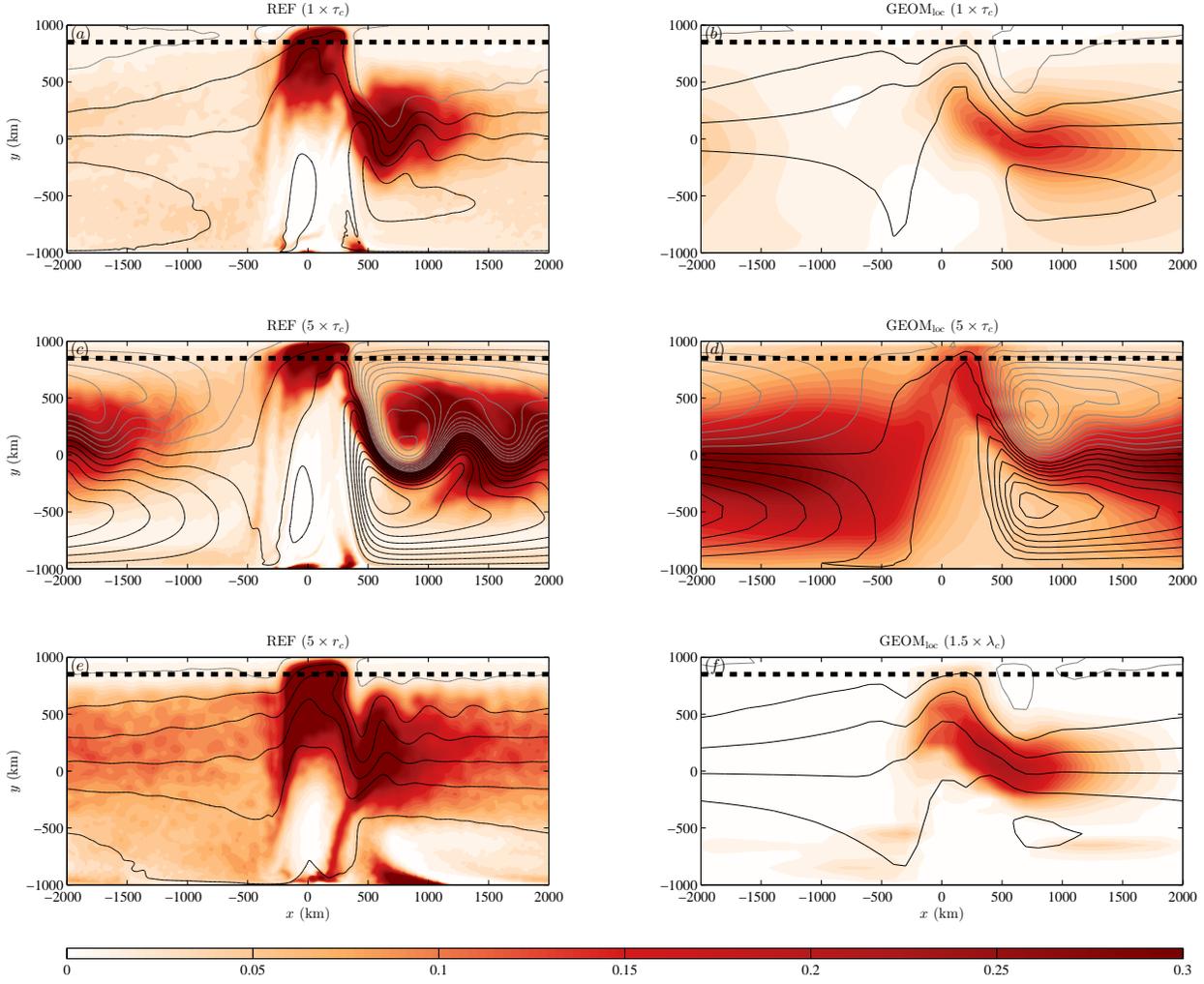

  \includegraphics[width=\linewidth]{{{channel_tave_eE_compare_full}}}
  \caption{Depth averaged eddy energy distribution for REF and for
  \mbox{GEOM$_{\rm loc}$} at: ($a,b$) control case at $(1\times \tau_c, 1\times
  r_c/\lambda_c) = (0.2\ \mathrm{N}\ \mathrm{m}^{-2}, 1.1\times10^{-3}\
  \mathrm{m}\ \mathrm{s}^{-1} / 10^{-7}\ \mathrm{s}^{-1})$; ($c,d$) large wind
  case at $(5\times \tau_c, 1\times r_c)$; ($e,f$) large dissipation case at
  $(1\times \tau_c, 5\times r_c / 1.5\times \lambda_c)$. Contours denote the
  Eulerian transport streamfunction (black: positive values starting at 25 Sv in
  spacings of 25 Sv; grey: negative values starting at 0 Sv in spacings of 25
  Sv). The dashed white line highlights the edge of the sponge region. The
  color scale is saturated, with limits chosen to demonstrate features between
  the calculations over a fixed color scale.}
  \label{fig:channel_tave_eE_compare}
\end{figure*}

For \mbox{GEOM$_{\rm int}$} and \mbox{GEOM$_{\rm loc}$}, the emergent $\langle E
\rangle$ increases with increasing wind stress, though not necessarily at the
same rate as REF. The rate of increase for \mbox{GEOM$_{\rm int}$} is slightly
sub-linear, as opposed to the predicted linear scaling given in
\cite{Mak-et-al17}. The increase in $\langle \kappa_{\rm gm} \rangle$ is also
slightly sub-linear, consistent with the behavior of $\langle E \rangle$. More
variation is shown in \mbox{GEOM$_{\rm loc}$} in both the resulting $\langle
E\rangle$ and $\langle \kappa_{\rm gm}\rangle$ levels, though the increase
roughly follows that of \mbox{GEOM$_{\rm int}$}. It should be noted that while
$\langle \kappa_{\rm gm} \rangle \leq \kappa_{\max}$ in \mbox{GEOM$_{\rm loc}$},
locally $\kappa_{\max}$ does get applied to the emergent $\kappa_{\rm gm}$
albeit in a small region of the domain (where the parameterized eddy energy is
large, see Figure~\ref{fig:channel_tave_eE_compare}$c,d$). At the lower peak
wind stress values, the emergent eddy energy value from \mbox{GEOM$_{\rm int}$}
and \mbox{GEOM$_{\rm loc}$} is much smaller than \mbox{REF}, which is consistent
with the circumpolar transport of \mbox{GEOM$_{\rm int}$} and \mbox{GEOM$_{\rm
loc}$} being larger than REF in Figure~\ref{fig:channel_tave_vary_diag}($a$) as
a result of the reduced emergent $\kappa_{\rm gm}$. For \mbox{CONST}, the
diagnosed $\langle E\rangle$ is roughly two orders of magnitude smaller, and so
appears almost on the axes in this plot with linear scales.

For changing dissipation, while the sensitivity of the emergent $\langle E
\rangle$ with changing bottom drag coefficient $r$ in REF is consistent with the
eddy calculation reported in \cite{Marshall-et-al17}, and the sensitivity of the
emergent $\langle E \rangle$ is consistent with the \mbox{GEOM$_{\rm int}$}
calculations reported in \cite{Mak-et-al17}, these display opposite sensitivity
to each other. The resulting sensitivity of $\langle \kappa_{\rm gm} \rangle$ in
the coarse resolution calculations is consistent with the decreasing $\langle E
\rangle$, as well as the circumpolar transport increasing, though the cause and
effect is more convoluted \citep[see discussion in][]{Mak-et-al17}. This
discrepancy indicates that the difference between changing $\lambda$ and $r$,
while broadly agreeing in other diagnostics, is more subtle in the resulting
eddy energetics. This discrepancy is discussed at the end of this article.

The advantage of \mbox{GEOM$_{\rm loc}$} over \mbox{GEOM$_{\rm int}$} is the
ability to provide a horizontal spatial structure through the emergent
depth-integrated eddy energy field. Figure~\ref{fig:channel_tave_eE_compare}
shows the depth-averaged eddy energy field, together with the transport
streamfunction of REF and \mbox{GEOM$_{\rm loc}$} for the control case, the
large wind case and the large dissipation case. The eddy energy is mostly
concentrated downstream of the ridge (located in the region $-400\ \ \mathrm{km}
\leq y \leq 400\ \ \mathrm{km}$). The eddy energy signature extends further with
increased wind. It is particularly noteworthy that the parameterized eddy energy
(which strongly correlates with the emergent $\kappa_{\rm gm}$) is able to
capture aspects of the eddy energy signature displayed by REF. The emergent
values may differ but additional tuning may be done to provide a better match of
the magnitude. An interesting observation is that the eddy energy is extended
too far to the east, which may be remedied by eddy energy propagation westward
at the long Rossby phase speed, a feature not included here. However, it is
noted that eddies are observed to propagate eastward at the long Rossby wave
phase speed within the core of the Antarctic Circumpolar Current
\citep{KlockerMarshall14}, so the effect of including the additional propagation
is unclear.

For completeness, the eddy energy field at the large dissipation value is also
included. At larger $r$, the dominant contribution of the eddy energy in REF
comes from the EPE. On the other hand, increasing $\lambda$ in \mbox{GEOM$_{\rm
loc}$} appears to instead concentrate the eddy energy around the ridge, with an
increase in the magnitude over the ridge. The signature pattern is not entirely
different to the control case and in fact resembles well the general EKE pattern
of REF (not shown). 


\section{Sector configuration}\label{sec:sector}

To compare the characteristics of \mbox{GEOM$_{\rm int}$} and \mbox{GEOM$_{\rm
loc}$} in a more complex setting, a sector configuration with a re-entrant
channel connected to an ocean basin is employed. Besides a circumpolar
transport, this configuration allows the possibility of a latitudinally extended
residual meridional overturning circulation (RMOC). A growing number of analyses
and results from eddy permitting numerical models suggests that while the
circumpolar transport is largely insensitive to changes in wind forcing, the
RMOC shows some sensitivity to changes in wind forcing
\citep[e.g.,][]{Hogg-et-al08, FarnetiDelworth10, Farneti-et-al10, FarnetiGent11,
GentDanabasoglu11, Meredith-et-al12, MorrisonHogg13, Munday-et-al13,
Farneti-et-al15}, i.e., numerical ocean models are expected to be largely eddy
saturated, and partially eddy compensated. A sector configuration allows for
study of whether the \mbox{GEOM$_{\rm int}$} and \mbox{GEOM$_{\rm loc}$} have
the potential in reproducing both eddy saturation and eddy compensation effects,
in a more complex and realistic setting.


\subsection{Setup}

The sector configuration detailed in \cite{Munday-et-al13} was employed. As a
brief summary, the domain spans $60^\circ$S to $60^\circ$N in latitude, with a
re-entrant channel from $60^\circ$S to $40^\circ$S, connected to a narrow basin
of $20^\circ$ in longitude. The model employs the \cite{JackettMcDougall95}
nonlinear equation of state. The depth is $5000\ \mathrm{m}$ everywhere except
for a $1^\circ$ wide ridge of height $2500\ \mathrm{m}$ located on the eastern
side of the channel (or 1 grid box for the $2^\circ$ coarse resolution
calculations), which blocks the $f/H$ contours (see Fig.~1 of
\citealt{Munday-et-al13}). An idealized wind forcing centered just north of the
channel of the form
\begin{equation}\label{eq:wind-sector}
  \tau_s = \begin{cases}
    \tau_0 \sin^2(\pi(y + 60) / 30), & \textnormal{if } y < -30,\\
    0, & \textnormal{otherwise},
  \end{cases}
\end{equation}
is imposed, where $\tau_0$ is the peak wind stress and $y$ is the latitude in
degrees. On the surface, the temperature and salinity is restored
to\footnote{Equation \eqref{eq:salinity-surf-sector} here corrects a typo in
equation 3 of \cite{Munday-et-al13}.}
\begin{equation}\label{eq:temp-surf-sector}
  T = \begin{cases}
    T_S + \Delta T \sin(\pi(y + 60) / 120), & y \geq 0,\\
    T_N + (\Delta T + T_S - T_N) \sin(\pi(y + 60) / 120), & y < 0,
  \end{cases}
\end{equation}
and
\begin{equation}\label{eq:salinity-surf-sector}
  S = \begin{cases}
    S_S + \Delta S (1 + \cos \pi y / 60) / 2, & y \geq 0,\\
    S_N + (\Delta S + S_S - S_N) (1 + \cos \pi y / 60) / 2, & y < 0,
  \end{cases}
\end{equation}
with $(T_S, T_N, \Delta T) = (0, 5, 30)^\circ \mathrm{C}$ and $(S_S, S_N, \Delta
S) = (34, 34, 3)\textnormal{psu}$, over a time-scale of $30$ and $10$ days
respectively. The vertical domain is discretized with $42$ uneven vertical
levels, thinnest of $10\ \mathrm{m}$ at the top, down to the thickest of $250\
\mathrm{m}$ at the bottom. All other details are as reported in
\cite{Munday-et-al13}.

In this instance, the eddy permitting reference calculation (REF) has a
$1/6^{\circ}$ horizontal grid spacing. The control simulation is taken to have
control peak wind stress $\tau_0 = \tau_c = 0.2\ \mathrm{N}\ \mathrm{m}^{-2}$
and control diapycnal diffusivity of $\kappa_d = \kappa_{d,c} = 3\times10^{-5}\
\mathrm{m}^2\ \mathrm{s}^{-1}$. The perturbation experiments at varying $\tau_0$
and now $\kappa_d$ (rather than $r$ in the channel calculations) were restarted
from perturbed states reported in \cite{Munday-et-al13} for a further $10$ years
for extra diagnostics (again, a model year is 360 days). The eddy permitting
reference calculations employ a biharmonic dissipation in the momentum equation
that maintains a grid scale Reynolds number to be 0.15. A spatially and
temporally constant GM eddy transfer coefficient of $\kappa_{\rm gm} = \kappa_0
= 0.26\ \mathrm{m}^2\ \mathrm{s}^{-1}$ is employed.

For the coarse resolution models, the horizontal spacing is $2^{\circ}$, with
the \mbox{CONST}, \mbox{GEOM$_{\rm int}$} and \mbox{GEOM$_{\rm loc}$} variants
considered; note that the domain of integration in \mbox{GEOM$_{\rm int}$} is
taken over the whole domain rather than, for example, just over the circumpolar
region. An initial calculation was first restarted from the $2^{\circ}$
simulation at the control parameter value of \cite{Munday-et-al13}, which is a
\mbox{CONST} calculation with $\kappa_0 = 1000\ \mathrm{m}^2\ \mathrm{s}^{-1}$,
for a further 1000 model years as a \mbox{CONST} calculation but with $\kappa_0
= 1500\ \mathrm{m}^2\ \mathrm{s}^{-1}$. Then perturbation experiments were
carried out for a further 1800 years, following by a time-averaging over a
further 200 years. The control linear eddy energy dissipation coefficient
$\lambda_c$ was chosen to be $\lambda_c = 10^{-7}\ \mathrm{s}^{-1}$, as in the
channel calculations. The values of $\alpha$ and $\kappa_0$ were tuned so that
the coarse resolution calculations have roughly the same emergent circumpolar
transport values as the control values for REF. These values were then fixed as
the wind forcing and dissipation parameters were varied. The relevant parameter
values are documented in Table~\ref{tbn:sector-params}.

\begin{table*}[tbp]
  \begin{center}
  {\small
    \begin{tabular}{|l|c|c|}
      \hline
      Parameter & Value & units\\
      \hline
      $\tau_0$ & 
        0.00, 0.01, 0.05, 0.10, 0.15, \underline{0.20}, 0.25, 0.30, 0.40, 0.60, 0.80, 1.00 &
        $\mathrm{N}\ \mathrm{m}^{-2}$\\
      \hline
      $\lambda$ & 
        1, 2, \underline{3}, 4, 5, 6, 8, 10, 14, 18, 22, 26, 30 &
        $10^{-5}\ \mathrm{m}^2\ \mathrm{s}^{-1}$\\
      \hline
      $\lambda$ & 
        0.80, 0.85, 0.90, 0.95, \underline{1.00}, 1.10, 1.20, 1.30, 1,40, 1.50 &
        $10^{-7}\ \mathrm{s}^{-1}$\\
      \hline
      $\kappa_0$ &
        1500 (\mbox{CONST}) &
        $\mathrm{m}^2\ \mathrm{s}^{-1}$\\
      \hline
      $\alpha$ &
        0.075 (\mbox{GEOM$_{\rm int}$}), 0.07 (\mbox{GEOM$_{\rm loc}$}) &
        ---\\
      \hline
    \end{tabular}
  }
  \end{center}
  \caption{Parameter values that were employed for the sector experiments. The
  values underlined are designated the control simulation.}
  \label{tbn:sector-params}
\end{table*}

Two diagnostics are computed to test first for the eddy saturation property;
again, the diagnostic quantities are time-averaged unless otherwize stated. The
total circumpolar transport is calculated as in \eqref{eq:transtotal}. Similar
to equation \eqref{eq:therm-depth}, a pycnocline depth diagnostic is obtained by
computing the pycnocline location quantity
\begin{equation}\label{eq:pycno-depth} 
  z_{\scriptsize \mbox{pyc}} = 2 \cfrac{\int_{-H}^0 z[\overline{\rho} - \overline{\rho}(z=-H)]\; \mathrm{d}z}{\int_{-H}^0 [\overline{\rho} - \overline{\rho}(z=-H)]\; \mathrm{d}z}, 
\end{equation} 
and averaging over the region between 30${}^\circ$S and 30${}^\circ$N, which
roughly gives an estimate of the pycnocline of the ocean, avoiding the southern
and northern regions where deep mixed layers may bias the results.


\subsection{Results}

\begin{figure}
  \includegraphics[width=\linewidth]{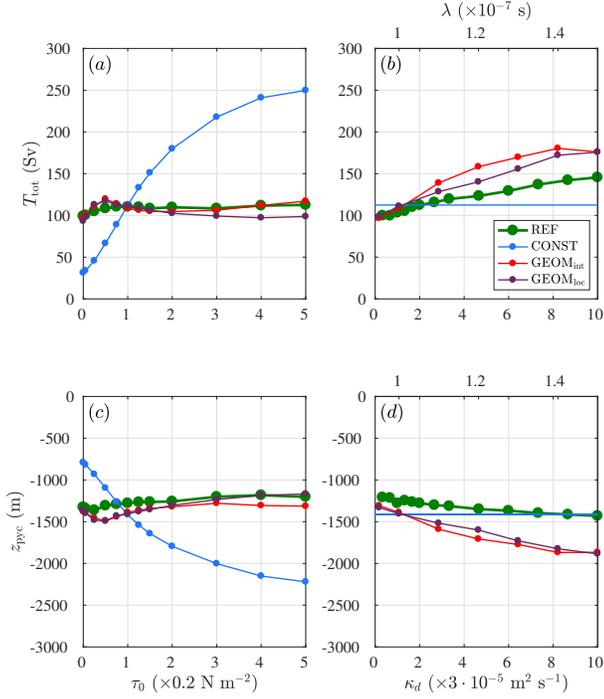}
  \caption{Diagnosed time-mean transports (in units of Sv) of reference and
  pycnocline location (in units of $\mathrm{m}$), for varying wind and diapycnal
  diffusivity, showing: ($a,b$) total circumpolar transport; ($c,d$) pycnocline
  depth of the basin.}
  \label{fig:sector_tave_vary_diag}
\end{figure}

Figure~\ref{fig:sector_tave_vary_diag} shows the aforementioned diagnostics at
varying wind and dissipation values. To summarize, for varying wind, the eddying
calculation REF possesses a circumpolar transport that displays weak dependence
on the peak wind stress and may be called eddy saturated. The pycnocline
location is also only weakly dependent on varying peak wind stress. For
increasing diapycnal diffusivity, the circumpolar transport increases and the
pycnocline depth increases (more negative $z_{\scriptsize \mbox{pyc}}$).
Assuming again that the circumpolar transport is dominated by thermal wind
transport and that isopycnals are essentially pinned at the outcropping regions,
increase in pycnocline depth are linked directly to increased circumpolar
transport via increasing the tilt of isopycnals.

With this in mind, \mbox{CONST} is categorically not eddy saturated, displaying
large sensitivity of the circumpolar transport and pycnocline depth to changing
wind forcing. On the other hand, both the circumpolar transport and pycnocline
location of \mbox{GEOM$_{\rm int}$} and \mbox{GEOM$_{\rm loc}$} display relative
insensitivity with changing peak wind stress, which is far more consistent with
the REF case. As in the channel configuration, $\lambda$ in the \mbox{GEOM$_{\rm
int}$} and \mbox{GEOM$_{\rm loc}$} calculations increases the circumpolar
transport and pycnocline depth, much like increasing $\kappa_d$ in REF.


\subsubsection{Varying wind experiments}

While \mbox{GEOM$_{\rm int}$} and \mbox{GEOM$_{\rm loc}$} are eddy saturated,
the associated sensitivity in the RMOC remains to be investigated. The RMOC may
be diagnosed via the MITgcm \verb|layers| package \citep{Abernathey-et-al11}.
The RMOC streamfunction is diagnosed as
\begin{equation}\label{eq:psi-rmoc}
  \Psi_{\mathrm{r}}(y, \rho) = -\int_{0}^{L_x} \int_{\rho_0}^\rho 
    \overline{hv}\; \mathrm{d}\rho'\; \mathrm{d}x,
\end{equation}
with $x$ is the longitude, $h = (\dy \rho/\dy z)^{-1}$ is the thickness, and the
time filter is carried out in density co-ordinates. For this sector model with a
nonlinear equation of state, the calculations are carried out in potential
density co-ordinates. The potential density $\rho$ is referenced to the
30${}^\textnormal{th}$ model level (at around $2000\ \mathrm{m}$ depth), and the
binning is over 241 discrete layers between $1031$ and $1037\ \mathrm{kg}\
\mathrm{m}^{-3}$, equally spaced at $0.025\ \mathrm{kg}\ \mathrm{m}^{-3}$. For a
simulation with an explicit representation of the eddy field, the RMOC
streamfunction encapsulates both contributions from the Eulerian overturning
circulation and eddy-induced transport. For all the simulations, since there is
a parameterized component of the circulation via parameterized eddy induced
velocity, the additional component needs to be added in (this is very weak in
REF since $\kappa_0$ was chosen to be very small).

\begin{figure*}
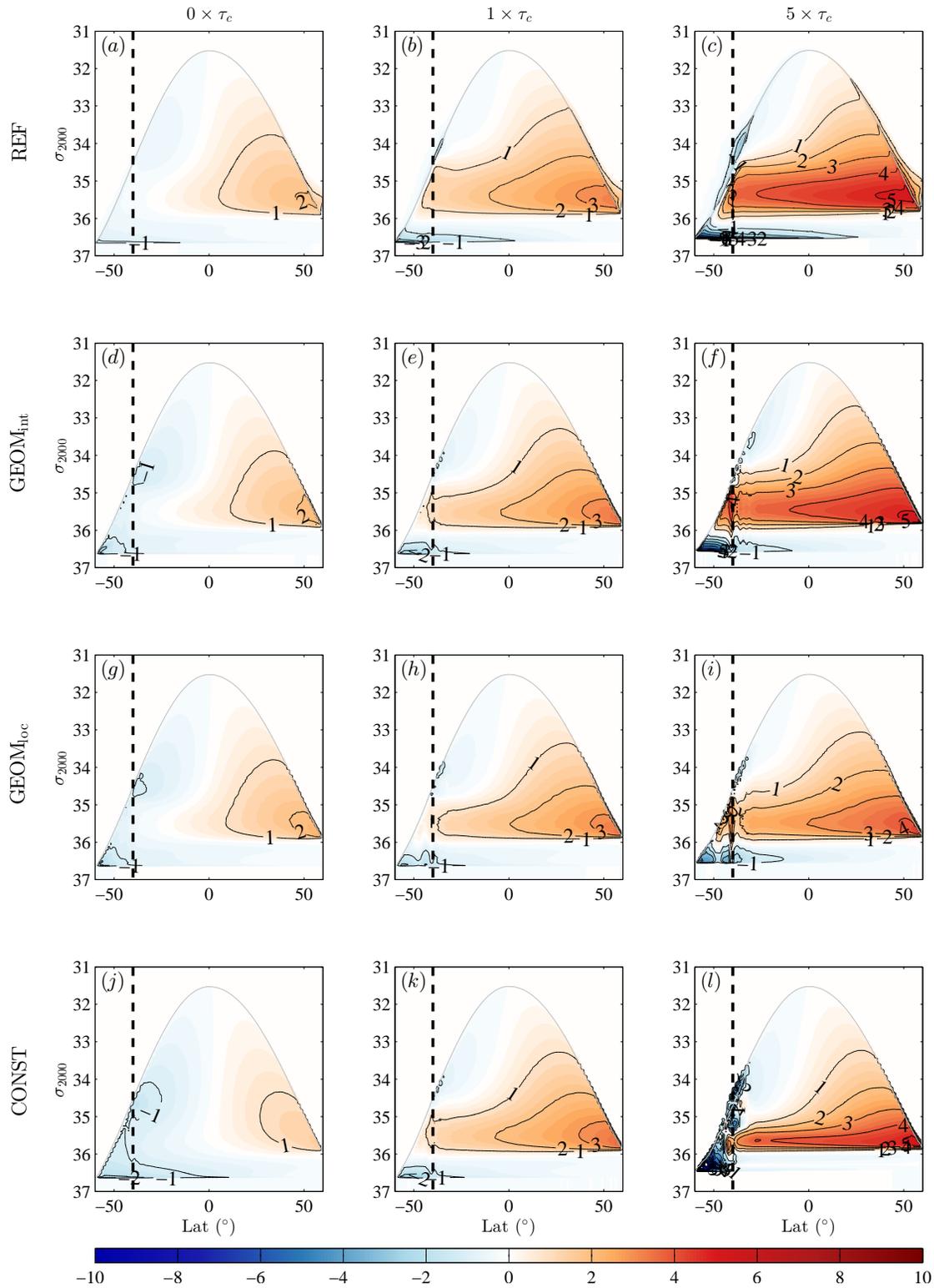

\begin{center}
  \includegraphics[width=0.9\linewidth]{{{sector_tave_rmoc_compare}}}
  \caption{RMOC streamfunction with the GM contribution (in units of
  $\mathrm{Sv}$) at varying wind forcing for REF ($a$--$c$), \mbox{GEOM$_{\rm
  int}$} ($d$--$f$), \mbox{GEOM$_{\rm loc}$} ($g$--$i$) and \mbox{CONST}
  ($j$--$l$). Shading and thin black contours are both contours of the
  streamfunction, at spacings of $\pm 0.25$ and $\pm 1\ \mathrm{Sv}$
  respectively (zero contour removed); red is clockwise circulation and blue is
  counter-clockwise circulation. The grey contour is the zonally averaged
  surface potential density contour. The dashed line indicates the edge of the
  re-entrant channel.}
  \label{fig:sector_tave_rmoc_compare}
\end{center}
\end{figure*}

The diagnosed RMOCs for varying wind forcing are shown in
Figure~\ref{fig:sector_tave_rmoc_compare}. Focusing first on the control case
for REF (Figure~\ref{fig:sector_tave_rmoc_compare}$b$; cf. Figure~8$c$ of
\citealt{Munday-et-al13}), it may be seen that the RMOC consists of two main
cells: (i) an upper positive cell that is the model analogue of the North
Atlantic Deep Water (NADW), downwelling in the northern hemisphere and upwelling
in the model Southern Ocean region; (ii) a lower negative cell that is the model
analogue of the Antarctic Bottom Water (AABW), controlled largely by the
convective activity occurring in the southern edges of the domain. Additionally,
there is an Antarctic Intermediate Water (AAIW) negative cell slightly north of
the NADW upwelling region, characterized by shallow convection. Excursions above
the surface potential density contour represents significant eddy density
transport giving rise to locally higher densities in time/space.

For the control wind forcing, the global morphology of the RMOC appears to be
well captured in all the coarse resolution calculations, as seen in
Figure~\ref{fig:sector_tave_rmoc_compare}($e,h,k$) for \mbox{GEOM$_{\rm int}$},
\mbox{GEOM$_{\rm loc}$} and \mbox{CONST} respectively. The main differences
arise in the excursion of the RMOC above the time-zonal-mean surface density in
the north and in the details of the AABW negative cell. The former is because
there are no explicit mesoscale eddies in the coarse resolution calculations.
The latter is likely much more subtle since this involves convective processes
responsible for the formation of AABW, as well as the vertical response of the
eddy field via supplying the warm, salty NADW water to be transformed into AABW,
and in setting the extent of the AABW cell via the eddy induced circulation.

When varying wind forcing, the changes in the RMOC displayed by REF are largely
matched by \mbox{GEOM$_{\rm int}$} and \mbox{GEOM$_{\rm loc}$}. At no wind
forcing, the NADW positive cell is approximately of the same magnitude and with
similar extents into the southern hemisphere. At large wind, the increase in
magnitude and extent in both the NADW positive cell and AABW negative cell are
seen. Both \mbox{GEOM$_{\rm int}$} and \mbox{GEOM$_{\rm loc}$} struggle to
reproduce the latitudinal extent and the strength of the AABW negative cell.
However, both \mbox{GEOM$_{\rm int}$} and \mbox{GEOM$_{\rm loc}$} certainly
appear to provide improvement on \mbox{CONST}, where the latitudinal extent of
the NADW at zero wind forcing differs significantly from REF, and increased
noise in the AABW cell and a NADW cell spanning over a smaller set of water mass
classes at large wind forcing. This enhanced level of noise in and just north of
the channel region in \mbox{CONST} coincides with increased convective activity
in the same regions, where the prescribed $\kappa_{\rm gm} = \kappa_0$ is
overwhelmed by the strong Eulerian overturning cell, leading to steep isopycnals
and increased convective activity that is absent in REF.


\subsubsection{Varying dissipation experiments}

\begin{figure}
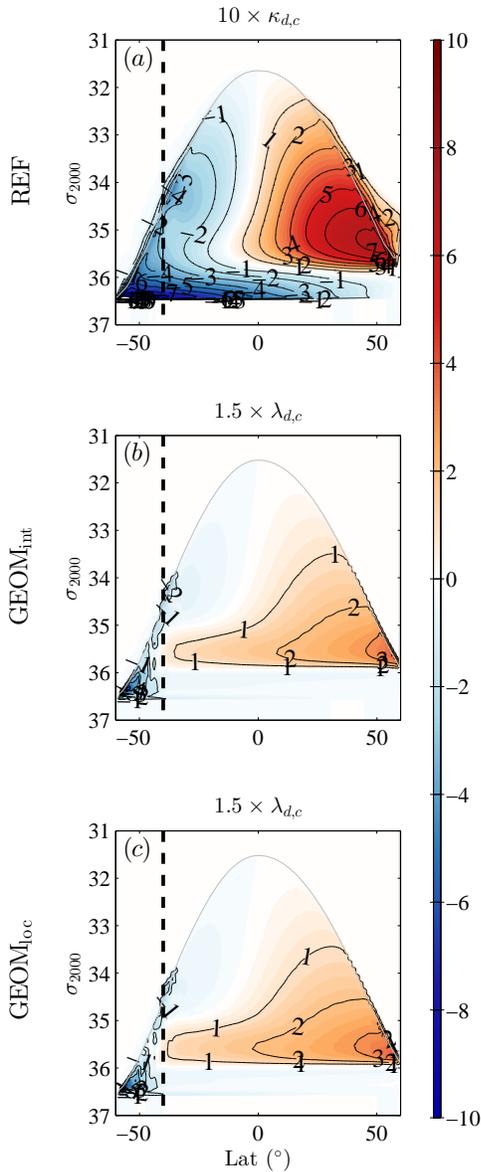

  \begin{center}
  \includegraphics[width=0.8\linewidth]{{{sector_tave_rmoc_compare_vary_diss}}}
  \end{center}
  \caption{RMOC streamfunction with the GM contribution (in units of
  $\mathrm{Sv}$) at large dissipation for REF ($a$, $10\times\kappa_{d,c}$),
  \mbox{GEOM$_{\rm int}$} ($b$, $1.5\times\lambda_c$) and \mbox{GEOM$_{\rm
  loc}$} ($c$, $1.5\times\lambda_c$) at control peak wind stress $\tau_c$.
  Shading and thin black contours are both contours of the streamfunction, at
  spacings of $\pm 0.25$ and $\pm 1\ \mathrm{Sv}$ respectively (zero contour
  removed); red is clockwise circulation and blue is counter-clockwise
  circulation. The grey contour is the zonally averaged surface potential
  density contour. The dashed line indicates the edge of the re-entrant
  channel.}
  \label{fig:sector_tave_rmoc_compare_vary_diss}
\end{figure}

Increasing diapycnal diffusivity $\kappa_d$ increases the rate of water mass
transformation, which deepens the pycnocline, thus leading to a larger region
with thermal wind transport, consistent with the diagnoses of results shown in
Figure~\ref{fig:sector_tave_vary_diag}($b,d$). While increasing $\lambda$ can
reproduce sensitivities in the circumpolar transport, this is not the case in
the resulting RMOC streamfunction, displayed in
Figure~\ref{fig:sector_tave_rmoc_compare_vary_diss} at the large dissipation
scenario ($10\times\kappa_{d,c}$ for \mbox{REF}, $1.5\times\lambda_c$ for
\mbox{GEOM$_{\rm int}$} and \mbox{GEOM$_{\rm loc}$}). At such a large $\kappa_d$
for REF, the increased rate of water mass transformation results in a RMOC with
a NADW positive cell that is latitudinally confined, since the water mass is
transformed within the basin before it can upwell in the re-entrant channel. On
the other hand, a latitudinally extended RMOC is still somewhat maintained at
large $\lambda$ for \mbox{GEOM$_{\rm int}$} and \mbox{GEOM$_{\rm loc}$}. The
appearance of noise in the AABW cell may be attributed to the fact that the
value of the emergent eddy energy and thus $\kappa_{\rm gm}$ has decreased (cf.
Figure~\ref{fig:sector_tave_vary_diag_eke_kappa}$b,d$), and an imbalance in the
eddy induced and Eulerian overturning leads to increased convective activity. 

\subsubsection{Other emergent quantities}

\begin{figure}
  \includegraphics[width=\linewidth]{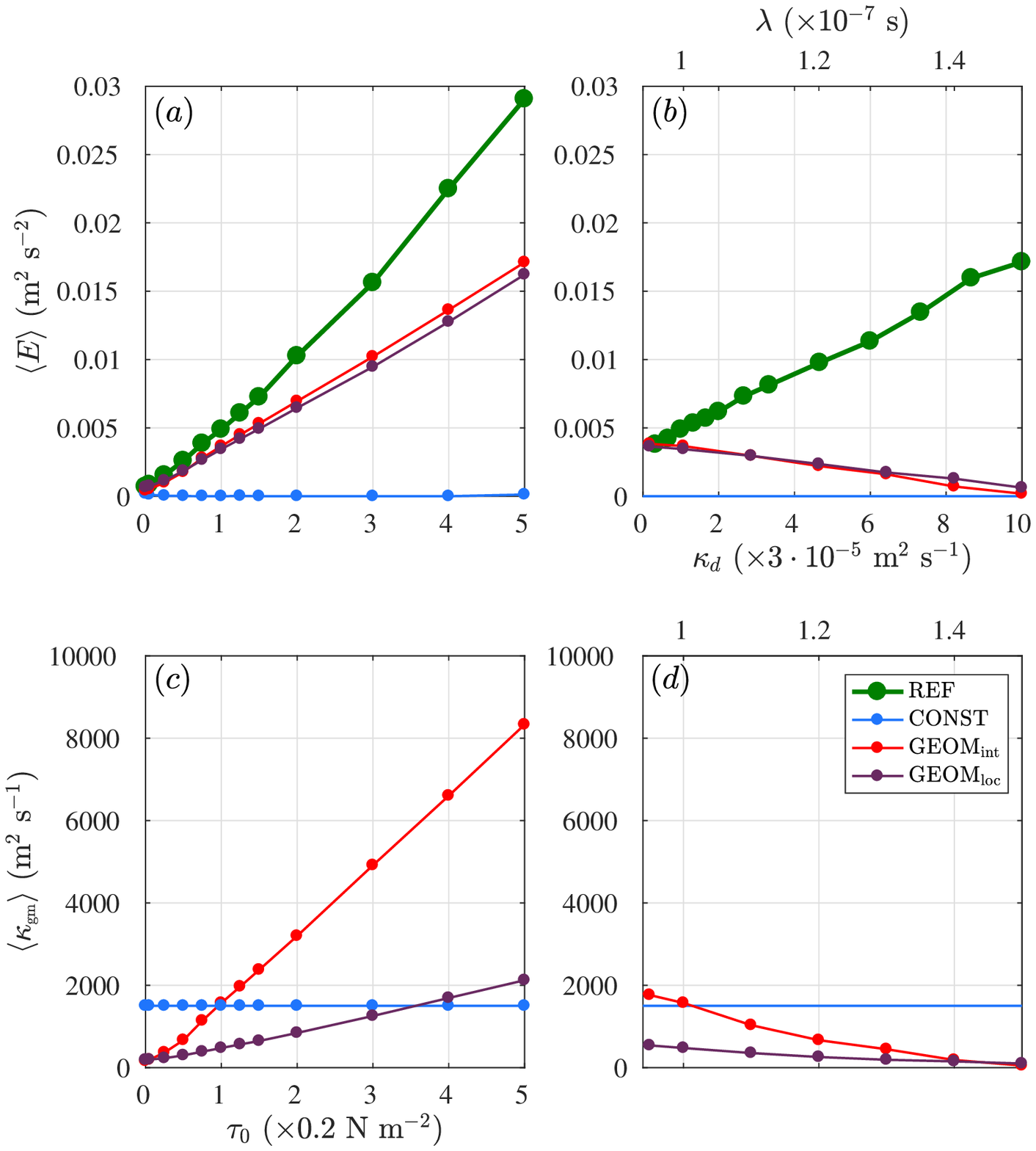}
  \caption{Diagnosed outputs relating to the paramaeterisation variants for the
  sector model, for varying wind ($a,c$) and varying dissipation ($b,d$),
  showing: ($a,b$) domain-averaged eddy energy in the (in units of
  $\mathrm{m}^2\ \mathrm{s}^{-2}$); ($c,d$) domain-averaged GM coefficient
  for parameterized models (in units of $\mathrm{m}^2\
  \mathrm{s}^{-1}$).}
  \label{fig:sector_tave_vary_diag_eke_kappa}
\end{figure}

\begin{figure*}
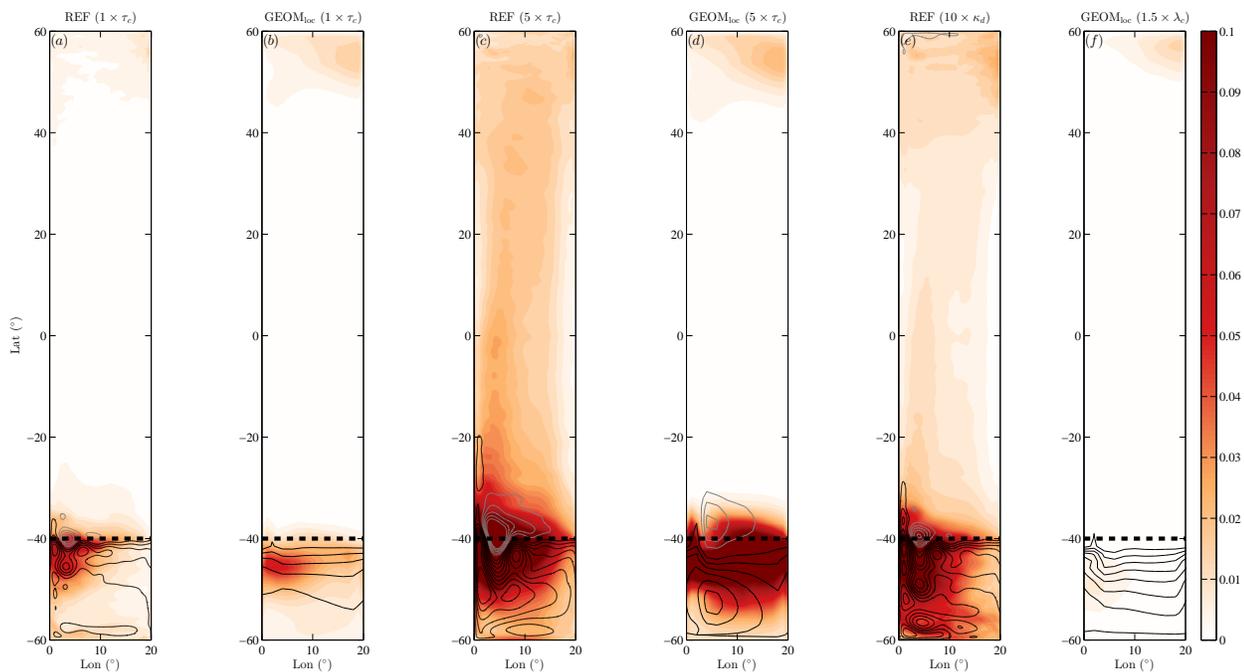

\begin{center}
  \includegraphics[width=\linewidth]{{{sector_tave_eE_compare_full}}}
  \caption{Depth-averaged total eddy energy for REF and \mbox{GEOM$_{\rm loc}$}
  at ($a,b$) control case at $(1\times \tau_c, 1\times r_c/\lambda_c) = (0.2\
  \mathrm{N}\ \mathrm{m}^{-2}, 1.1\times10^{-3}\ \mathrm{m}\ \mathrm{s}^{-1} /
  10^{-7}\ \mathrm{s}^{-1})$; ($c,d$) large wind case at $(5\times \tau_c,
  1\times \kappa_{d,c}/\lambda_c)$; ($e,f$) large dissipation case at $(1\times
  \tau_c, 10\times \kappa_{d,c} / 1.5\times \lambda_c)$. Contours denote the
  Eulerian transport streamfunction (black: positive values starting at 0 Sv in
  spacings of 20 Sv; grey: negative values starting at -100 Sv in spacings of 20
  Sv). The dashed white line highlights the edge of the reentrant channel. The
  color scale is saturated, with limits chosen to demonstrate features between
  the calculations over a fixed color scale.}
  \label{fig:sector_tave_eE_compare}
\end{center}
\end{figure*}

Figure~\ref{fig:sector_tave_vary_diag_eke_kappa} shows the domain-averaged eddy
energy $\langle E\rangle$ and domain-averaged GM eddy transfer coefficient
$\langle \kappa_{\rm gm} \rangle$ for varying input parameters, diagnosed as in
the channel configuration (now with potential density instead of temperature as
the gridding field when using the \verb|layers| package). In this particular
instance, the diagnosed domain-averaged value of EKE and EPE for REF is roughly
equal in magnitude at control peak wind stress, but EKE becomes dominant
especially in the channel region at large wind stress. For \mbox{GEOM$_{\rm
int}$} and \mbox{GEOM$_{\rm loc}$}, $\langle E\rangle$ increases with increasing
wind, at a roughly linear rate, which is consistent with the prediction given in
\cite{Mak-et-al17}. Note that $\langle E\rangle$ for REF is increasingly
super-linearly (cf., \cite{Munday-et-al13} but for the domain-averaged EKE). The
diagnosed $\langle E\rangle$ for \mbox{CONST} is typically two order of
magnitudes smaller and appears on the axes in this plot with linear scales. The
increase in $\langle \kappa_{\rm gm}\rangle$ for \mbox{GEOM$_{\rm int}$} and
\mbox{GEOM$_{\rm loc}$} is consistent with the increase in eddy energy. The
emergent $\langle \kappa_{\rm gm} \rangle$ for \mbox{GEOM$_{\rm loc}$} is
smaller since $\kappa_{\rm gm}$ is small over the basin; locally in the channel
however $\kappa_{\rm gm}$ can be large via a large local eddy energy, and
$\kappa_{\max}$ takes over in the model Antarctic Circumpolar Current (cf.
Figure~\ref{fig:sector_tave_eE_compare}$c,d$).

With increasing dissipation, again there is further suggestion that while
increasing $\lambda$ results in decreased $\langle E\rangle$ in \mbox{GEOM$_{\rm
int}$} and \mbox{GEOM$_{\rm loc}$}, consistent with the findings of the channel
configuration and the results in \cite{Mak-et-al17}, it does not capture the
changes displayed by changing $\kappa_d$ in REF.

Finally, Figure~\ref{fig:sector_tave_eE_compare} shows the emergent
depth-averaged total eddy energy field and the transport streamfunction for
\mbox{REF} and \mbox{GEOM$_{\rm loc}$} at the control case, the large wind case,
and the large dissipation case. For REF, EKE and EPE contributions to the total
eddy energy are roughly equal, with the EKE contribution going from around 20\%
at zero wind to 80\% at the largest wind forcing, and staying around 60\% for
changing diapycnal diffusion. Comparing with \mbox{GEOM$_{\rm int}$}, observe
that, like the channel setting, the pattern of the emergent parameterized eddy
energy --- again strongly correlating with the emergent $\kappa_{\rm gm}$ ---
resembles the diagnosed total eddy energy from REF around the channel region,
and also in the northern hemisphere downwelling region at the control case and
large wind case. Both regions possess steep isopycnals outcropping at the
surface from the imposed surface restoring conditions, allowing for eddy energy
to grow. For both REF and \mbox{GEOM$_{\rm loc}$}, the eddy energy is large on
the western part of the channel, decreasing to the east. Again, the emergent
parameterized eddy energy in \mbox{GEOM$_{\rm loc}$} is more extended to the
east than \mbox{REF}, which may again be remedied by including the eddy energy
advection at the long Rossby phase speed.

At large wind forcing, the \mbox{REF} calculation displays substantially larger
eddy energy values even within the basin compared to \mbox{GEOM$_{\rm loc}$}. At
eddy permitting resolutions, eddies generated from the channel as well as the
northern sinking region may travel into the basin that, together with the
presence of waves, will contribute to the eddy energy signature seen in
\mbox{REF}. While it is reassuring to see that \mbox{GEOM$_{\rm loc}$} is
performing in the regions where baroclinic instability is expected to strong, it
remains a theoretical and modelling challenge to represent such advective
effects.

Increasing diapycnal diffusion results in a larger eddy energy signature in the
basin. These changes however are not captured by \mbox{GEOM$_{\rm loc}$} when
increasing $\lambda$, again demonstrating the discrepancy between the two
dissipations.


\section{Discussion and concluding remarks}\label{sec:conc}

This article has outlined and described the implementation of GEOMETRIC
(``Geometry of Ocean Mesoscale Eddies and Their Rectified Impact on Climate'')
in a three dimensional primitive equation ocean model. The GEOMETRIC recipe
utilizes the Gent--McWilliams formulation but with the eddy transfer coefficient
$\kappa_{\rm gm} = \alpha E (N / M^2)$, derived through rigorous mathematical
bounds \citep{Marshall-et-al12}, and with a linear dependence on the total eddy
energy. This is coupled to a parameterized budget for the depth-integrated total
eddy energy budget \citep[cf.][]{EdenGreatbatch08}. Done this way, the
parameterization of mesoscale eddies is still through an induced adiabatic
stirring as in the Gent--McWilliams scheme, but becomes energetically
constrained in the vertical and varies in the horizontal through the emergent
eddy energy signature. The coarse resolution calculations utilising variants of
the GEOMETRIC parameterization presented here are able to capture the bulk model
sensitivities of corresponding reference calculations with an explicit mesoscale
eddy field. In particular, for varying wind forcing, the coarse resolution
sector model employing GEOMETRIC is eddy saturated and, furthermore, the
resulting residual meridional overturning circulation also bears remarkable
resemblance to the eddy permitting reference calculation, showing potential for
reproducing eddy compensation.

On the other hand, this work has highlighted several subtleties, in particular
with respect to eddy energy dissipation, that need to be addressed. The
following discussion will focus on details of the parameterization, but it is
recognized that, for example, other model details such as bathymetry play a
central role in shaping the RMOC \citep[e.g.,][]{HoggMunday14, Ferrari-et-al16,
DeLavergne-et-al17} and will also affect the overall model response.

While the calculations with GEOMETRIC appear to capture the bulk morphological
changes of the RMOC over changing wind forcing, there are features that are at
odds with the reference calculation, notably in the strength and extent of the
model AABW. A candidate in improving the emergent RMOC response is to
incorporate a vertically varying eddy response. While this article presents
results for a vertically uniform eddy transfer coefficient ($\Gamma(z) \equiv
1$), it has long been recognized that the eddy transfer coefficient should vary
in the vertical \citep[e.g.,][]{Ferreira-et-al05}. Further, since the eddy
activity is expected to be strongest near the surface, the treatment of the
mesoscale parameterization scheme near the ocean surface is likely going to have
a large impact on the model response \citep[e.g.,][]{Danabasoglu-et-al08}.

A set of calculations with the structure function $\Gamma(z) = N^2 / N^2_{\rm
ref}$ \citep{Ferreira-et-al05} was carried out. While the associated coarse
resolution calculations following the GEOMETRIC prescription captures the
sensitivity in the circumpolar transport (and, in particular, is eddy saturated
in the sector model), care needs to be taken so other model aspects are also
reproduced. For example, in the sector configuration, an initial set of
calculations with $0 \leq \Gamma(z) \leq 1$ results in a shutdown of the
latitudinally extended RMOC. The reason for this is that the resulting eddy
response, while surface intensified, shuts off near the interface between the
channel and the basin, and thus the Eulerian overturning acts unopposed in that
region, causing the basin stratification to change substantially. Sample
calculations with a larger imposed $\kappa_{\min}$ and/or a lower bound on
$\Gamma(z)$ (e.g. $\Gamma_{\min} = 0.1$ as in \citealt{DanabasogluMarshall07})
result in a latitudinally extended RMOC. Other choices of vertical structure are
possible \citep[e.g.,][]{Ferrari-et-al08, Ferrari-et-al10}, which may be coupled
to mixed layer schemes \citep[e.g.,][]{Large-et-al94} and/or slope tapering
schemes \citep[e.g.,][]{Gerdes-et-al91}, all introducing additional tuning
parameters. In summary, comprehensive investigation of the RMOC response under
GEOMETRIC requires careful considerations of the vertical variation of the eddy
transfer coefficient, among other modelling details, and is deferred to a future
study.

The theory behind GEOMETRIC addresses the slumping of density surfaces in
baroclinic instability. While isopycnal slumping and eddy induced stirring (from
\citealt{GentMcWilliams90} and \citealt{Redi82} respectively) are often
implemented together \citep[e.g.,][]{Griffies98, Griffies-et-al98}, in this work
$\kappa_{\rm redi}$ was fixed to be a constant in space and time, while
$\kappa_{\rm gm}$ follows the GEOMETRIC prescription. Changing $\kappa_{\rm
redi}$ is expected to affect tracer transport and is thus of great importance in
the study of the ocean's role in heat transport and carbon storage, to name a
few \citep[e.g.,][]{PradalGnanadesikan14, AbernatheyFerreira15}. This is beyond
the scope of this work. It is noted here that diagnoses of isopycnal mixing in
numerical simulations appear to show $\kappa_{\rm redi}$ to be varying
vertically and depending linearly on the eddy energy
\citep[e.g.,][]{Abernathey-et-al13, AbernatheyFerreira15}. Analogous treatment
of $\kappa_{\rm redi}$ as outlined in this article may well be appropriate.

As discussed in the text, while eddy saturation is not expected to depend to
leading order on the lateral redistribution of eddy energy \citep{Mak-et-al17},
other details may. In the present implementation of GEOMETRIC, eddy energy is
advected by the depth-mean flow only, and the emergent eddy energy signature was
generally found to have a more eastward extension in the coarse resolution model
than the corresponding eddy permitting calculation. Inclusion of a westward
advective contribution at the long Rossby phase speed (consistent with
\citealt{Chelton-et-al07, Chelton-et-al11}, \citealt{Zhai-et-al10} and
\citealt{KlockerMarshall14}) is likely a remedy for the overly eastward
extension of the eddy energy signature. Taking the linear eddy energy damping
rate employed here at $10^{-7}\ \mathrm{s}^{-1}$ (dissipation rate inferred for
Southern Ocean from Zhai \& Marshall, pers. comm.) and a propagation speed of
$0.02\ \mathrm{m}\ \mathrm{s}^{-1}$, the extent of the energy signature is on
the order of $200\ \mathrm{km}$, which is approximately $2^{\circ}$ in
longitude. So while this effect may not be so significant in the Southern Ocean,
it is likely significant for western boundary currents, since the inferred
dissipation rate is lower in basins (Zhai \& Marshall, pers. comm.). The
inclusion and investigation into representing westward propagation by mesoscale
eddies is a subject of a future investigation.

Perhaps the most poorly constrained aspect of the present implementation of
GEOMETRIC is the treatment of eddy energy dissipation. Dissipation of mesoscale
eddy energy can be through a myriad of processes, such as bottom drag
\citep[e.g.,][]{Sen-et-al08}, lee wave radiation
\citep[e.g.,][]{NaveiraGarabato-et-al04, NikurashinFerrari11, Melet-et-al15},
western boundary processes \citep{Zhai-et-al10}, loss of balance
\citep[e.g.,][]{Molemaker-et-al05}, all of which vary in time, space, and
magnitude. Given the overwhelming complexity and the uncertainty in representing
such energy pathways, the choice of linear damping of eddy energy at a constant
rate over space was chosen to represent the collective effect of the
aforementioned processes in this first study of GEOMETRIC. With this choice, it
was found that coarse resolution models with GEOMETRIC are able to reproduce
sensitivities of the circumpolar transport and thermo/pycnocline depths of the
eddy permitting reference at varying wind forcing and dissipation. On the other
hand, the sensitivity of the domain-averaged eddy energy magnitude, while
reasonable for varying wind forcings, is completely at odds in the varying
dissipation experiments. Further investigation is required to reproduce the eddy
energetic sensitivities displayed in eddy permitting reference calculations. It
is anticipated here that a ``book-keeping'' approach, accounting for the energy
pathways through explicitly represented or parameterized components of the
dynamics in an ocean model, will prove to be the most fruitful approach. In the
present GEOMETRIC implementation, release of potential energy is accounted for
in the associated eddy energy budget, but one could envisage that a
parameterization for lee wave generation from geostrophic motions
\citep[e.g.,][]{Melet-et-al15} could serve as a spatially (and temporally)
varying sink in the GEOMETRIC eddy energy budget, or the sink of energy employed
in GEOMETRIC could be accounted for as a source in an energetically constrained
turbulence closure scheme \citep[e.g.,][]{Gaspar-et-al90, Madec-et-al98}, and so
forth. This proposed approach, alongside individual investigations of the
individual processes leading to energy transfer between scales is well beyond
the scope of this work here and is deferred to future investigations.


In closing, with the understanding that there are details that can be improved
upon, the results of this work lend further support to the GEOMETRIC framework
as a viable parameterization scheme that better parameterizes mesoscale eddies
in coarse resolution models, such that the resulting response in the emergent
mean matches more closely to models that explicitly represent mesoscale eddies.
For implementation into a global circulation ocean model, the primary change
required is to couple a depth-integrated eddy energy budget to the existing
Gent--McWilliams module. Diagnoses of eddy energetics via observations (e.g.,
Zhai \& Marshall, pers. comm.), idealized turbulence models \citep{Grooms15,
Grooms17} as well as ocean relevant simulations \citep[e.g.,][]{Stewart-et-al15}
will provide a first constraint on how to improve the representation of the
advection and dissipation of eddy energy, aiding in a more accurate and useful
representation of the ocean climatological response. In terms of approach, the
GEOMETRIC framework marks a shift of paradigm, from a focus on how to
parameterize eddy fluxes to focusing on parameterizing the eddy energetics and
the associated energy pathways.

\acknowledgments
This work was funded by the UK Natural Environment Research Council grant
NE/L005166/1 and NE/R000999/1 and utilized the ARCHER UK National Supercomputing
Service (http://www.archer.ac.uk). The lead author thanks Gurvan Madec for
discussions on implementation and energetic aspects of GEOMETRIC. The data used
for generating the plots in this article is available through the Edinburgh
DataShare service at \verb!http://dx.doi.org/10.7488/ds/2297!.








 \bibliographystyle{ametsoc_2014}
 \bibliography{refs}



\end{document}